\documentclass[conference]{IEEEtran}
\IEEEoverridecommandlockouts

\usepackage{cite}
\usepackage{amsmath,amssymb,amsfonts}
\usepackage{algorithmic}
\usepackage{graphicx}
\usepackage{url}
\usepackage{textcomp}
\usepackage{booktabs}       
\usepackage{multirow} 
\usepackage{cite}
\usepackage{xcolor}

\def\BibTeX{{\rm B\kern-.05em{\sc i\kern-.025em b}\kern-.08em
    T\kern-.1667em\lower.7ex\hbox{E}\kern-.125emX}}
    
\begin{document}

\title{Less is More: Data Curation Matters in Scaling Speech Enhancement}



\author{
\IEEEauthorblockN{
$^1$Chenda Li,\thanks{
This work was supported in part by the National Key Research and Development Program of China under Grant 2024YFC2418303, in part by the Shanghai Municipal Science and Technology Commission Project under Grant 2021SHZDZX0102, and in part by the Key Research and Development Program of Jiangsu Province, China (Grant No. BE2022059-4).}
$^1$Wangyou Zhang,
$^1$Wei Wang,
$^2$Robin Scheibler\thanks{Robin Scheibler from Google DeepMind served in an advisory role.},
$^3$Kohei Saijo,
$^4$Samuele Cornell,
$^5$Yihui Fu, \\
$^5$Marvin Sach,
$^6$Zhaoheng Ni, 
$^2$Anurag Kumar,
$^5$Tim Fingscheidt,
$^4$Shinji Watanabe,
$^1$Yanmin Qian
}
\IEEEauthorblockA{ 
\textit{$^1$Shanghai Jiao Tong University, China;
$^2$Google DeepMind, Japan;
$^3$Waseda University, Japan;}
}
\IEEEauthorblockA{
\textit{$^4$Carnegie Mellon University, USA;
$^5$Technische Universität Braunschweig, Germany;
$^6$Meta, USA}
}
}

\maketitle

\begin{abstract}

The vast majority of modern speech enhancement systems rely on data-driven neural network models. Conventionally, larger datasets are presumed to yield superior model performance, an observation empirically validated across numerous tasks in other domains. However, recent studies reveal diminishing returns when scaling speech enhancement data. We focus on a critical factor: prevalent quality issues in ``clean'' training labels within large-scale datasets. This work re-examines this phenomenon and demonstrates that, within large-scale training sets, prioritizing high-quality training data is more important than merely expanding the data volume. Experimental findings suggest that models trained on a carefully curated subset of 700 hours can outperform models trained on the 2,500-hour full dataset. This outcome highlights the crucial role of data curation in scaling speech enhancement systems effectively.

\end{abstract}


\section{Introduction}

Speech enhancement (SE)~\cite{loizouSpeechEnhancementTheory2007} remains a cornerstone of robust speech processing systems, enabling critical applications from hearing aids to voice communication devices. Its core objective is to improve speech quality and intelligibility by removing undesired acoustic components from degraded signals.
Thanks to the power of deep learning, neural network (NN)-based methods, including discriminative~\cite{xuRegressionApproachSpeech2015b,wangSupervisedSpeechSeparation2018a,luoConvTasNetSurpassingIdeal2019,huDCCRNDeepComplex2020,dubeyICASSP2023Deep2024,uni-versa-ext} and generative~\cite{luConditionalDiffusionProbabilistic2022,liDiffusionbasedGenerativeModeling2024,richterSpeechEnhancementDereverberation2023,leeFlowSEFlowMatchingbased2025} approaches, have made significant advances in speech enhancement over the past decade.

Due to the nature of data-driven supervised learning, training data critically governs system performance.
Conventional NN-based SE research has focused predominantly on narrowly defined tasks (e.g., additive noise suppression or dereverberation in isolation), typically employing small-scale, domain-specific datasets like VoiceBank+DEMAND~\cite{valentini-botinhaoSpeechEnhancementNoiseRobust2016}. 
 These datasets are often synthetically generated to match controlled evaluation conditions, which inherently limits their generalizability to complex, real-world environments where multiple distortions (e.g. noise and reverberation) coexist.

Increasing the quantity and richness of training data is a straightforward approach to improve the generalization ability of SE models.
A recent work makes a comprehensive study~\cite{zhangPerformancePlateausComprehensive2024} on the scalability of various discriminative speech enhancement models.
It is observed that the performance of discriminative SE could improve as the model size, model complexity, and dataset size increase, but tends to saturate when scaling beyond 157 hours of training data.
Meanwhile, another work \cite{gonzalezEffectTrainingDataset2024a} further investigates the effect on training data size (3 hours to 300 hours) of state-of-the-art discriminative and generative SE models.
The authors found that the performance improvement of the discriminative model was significant from 3 to 100 hours, but was relatively limited when scaled from 100 to 300 hours.
For generative models, the performance differences between systems trained for more than 10 hours are very minor.

On the other hand, to better explore and extend the generalization ability of speech enhancement models, the URGENT speech enhancement challenges \cite{zhangURGENTChallengeUniversality2024,koheiurgent2025} have encouraged researchers to use more diverse and larger-scale datasets to build universal and generalizable SE systems. 
In URGENT2025, the scope of speech enhancement has been extended to more complex tasks, including noise suppression, dereverberation, packet loss concealment, bandwidth extension, codec loss repair, etc \cite{koheiurgent2025}.
In terms of data amount, the URGENT2025 included more than 2,500 and 60,000 hours of speech source in \textit{Track1} and \textit{Track2}, respectively, to encourage participants to examine the impact of data amount. 
While deep learning has dominated the field of SE, the data-driven paradigm follows an implicit scaling hypothesis: larger datasets and increasingly complex models should yield consistent performance gains.
Surprisingly, there is no system in \textit{Track2} that consistently outperforms the best system in \textit{Track1}, which only utilizes a subset of the training data from \textit{Track2}. This suggests that simply increasing the data scale may not lead to effective performance improvement.

\begin{figure*}[!htb]
\centerline{\includegraphics[width=0.90\linewidth]{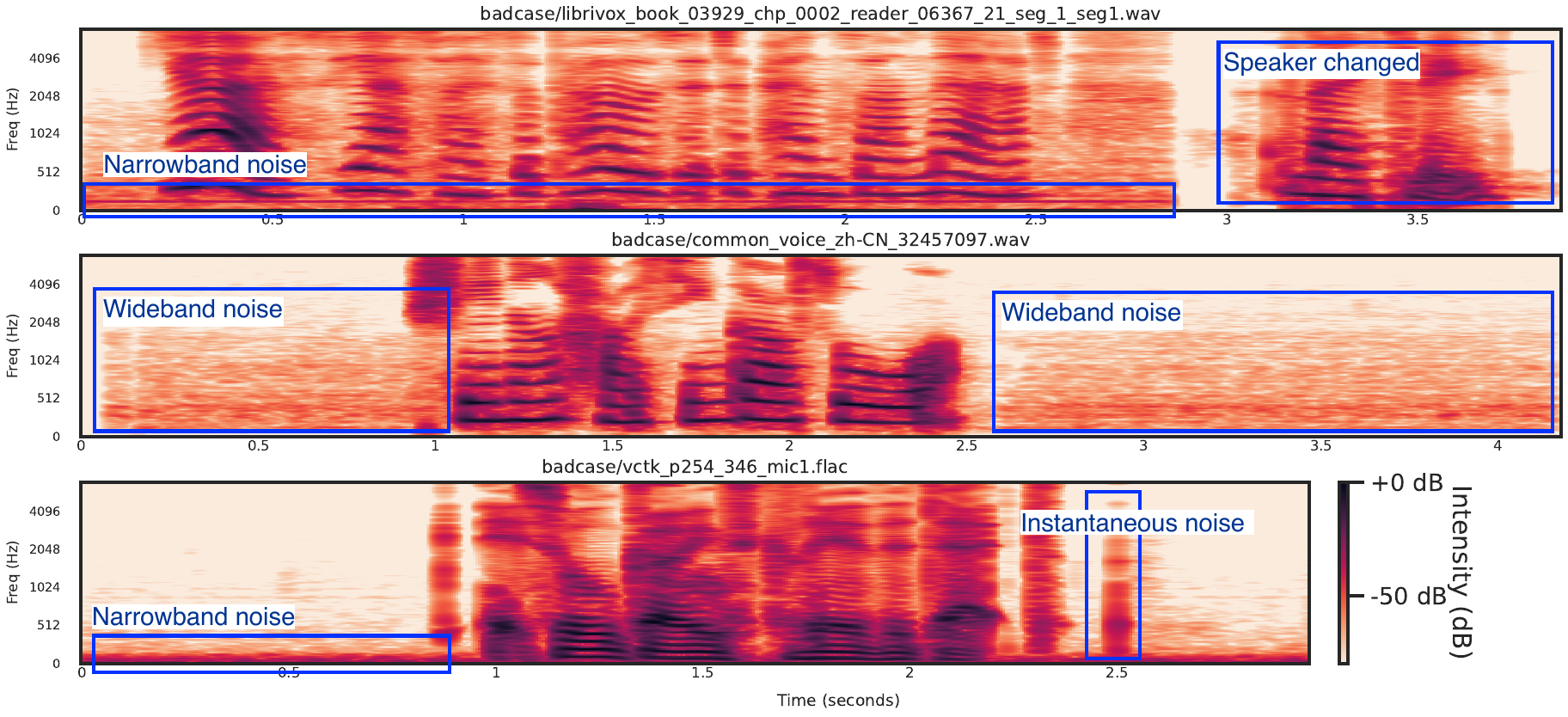}}
\caption{Training target of low quality selected from Librivox \cite{kearns2014librivox}, Common Voice \cite{ardilaCommonVoiceMassivelyMultilingual2020}, and VCKT \cite{veauxVoiceBankCorpus2013} datasets, which are commonly used in speech enhancement training \cite{dubeyICASSP2023Deep2024,luConditionalDiffusionProbabilistic2022,zhangURGENTChallengeUniversality2024,koheiurgent2025}. Audios are resampled to 22,050 Hz and displayed in log-scaled mel-spectrograms.}
\label{fig:badcase}
\end{figure*}

Based on the findings of the prior works~\cite{zhangPerformancePlateausComprehensive2024,gonzalezEffectTrainingDataset2024a,koheiurgent2025}, this paper revisits the role of data scaling in speech enhancement from a new perspective.
Acknowledging the known presence of low-quality samples within the ``clean'' labels of the URGENT2025 dataset (a key aspect of the challenge focused on noisy data utilization)~\cite{Lessons-Zhang2025}, we first perform a preliminary analysis on its 2,500-hour training set. Given this context, our study focuses on investigating the trade-off between data scale and quality.
We assume that these low-quality ``clean'' speech samples would have affected the scaling of the training data.
To verify this hypothesis, we apply multiple non-intrusive evaluation metrics to the 2500-hour speech source of the URGENT2025 training set, and select the top 100-hour, 350-hour, and 700-hour subsets.
We reviewed the effect of training data size on the performance of speech enhancement and compared it with the uniform random subset selection. 
Experiments in both discriminative and generative models show that curated subsets consistently outperform uniformly random samples of the same size.
Crucially, we discovered that the system trained on a 700-hour optimal subset outperforms the one trained on the full 2,500-hour set in multiple evaluation metrics.
This paper outlines our methodology and findings and proposes a data-centric data curation approach to advance future research on speech enhancement.

The rest of the paper is organized as follows: Section \ref{sec:exp_design} provides some preliminary analysis of the URGENT2025 training data, explains our motivation, and introduces our methods for data curation and experimental design.
We then detail our experimental setup and analyze the experimental results in Section~\ref{sec:exp_results}. Finally, Section~\ref{sec:conclusion} concludes this paper.

\section{Experimental Design}
\label{sec:exp_design}

\subsection{Task Definition}

In this work, we formulate the SE task as:
\begin{align}
	\hat{\mathbf{x}} &= \operatorname{SE}(\mathcal{F}(\mathbf{x})) \,, \label{eq:task}
\end{align}
where $\mathbf{x}$ is the clean speech signal and $\hat{\mathbf{x}}$ is the enhance speech signal.
$\operatorname{SE}$ is the speech enhancement system, and $\mathcal{F}(\cdot)$ is the distortion model that degrades the clean speech signal.
In most conventional SE works, the distortion model $\mathcal{F}(\cdot)$ typically considers only additive noise or reverberation in isolation.
We follow the extended SE task in the URGENT challenge \cite{zhangURGENTChallengeUniversality2024,koheiurgent2025}, where $\mathcal{F}(\cdot)$ includes seven different distortions:
1) additive noise, 2) reverberation, 3) clipping, 4) bandwidth limitation, 5) codec loss, 6) packet loss, and 7) wind noise.

\begin{figure*}[htb]
\centerline{\includegraphics[width=0.95\linewidth]{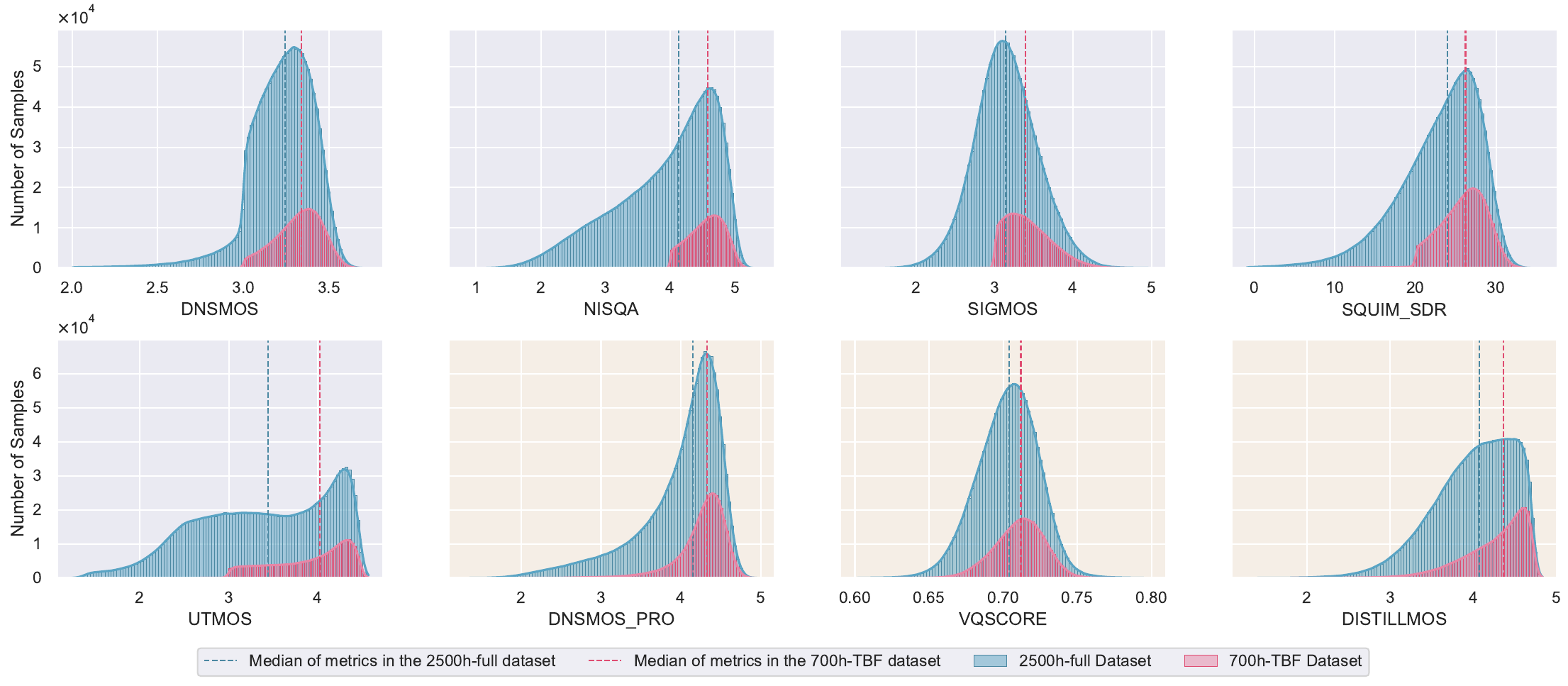}}
\caption{Histogram of non-intrusive metrics of the target speech source on the full dataset and Threshold-Based-Filtered (TBF) dataset. It is noted that the DNSMOS Pro \cite{cumlinDNSMOSProReducedSize2024}, VQScore \cite{fuSelfSupervisedSpeechQuality2023}, and Distill-MOS \cite{stahlDistillationPruningScalable2025} are not used as a threshold in TBF. After TBF, the median of all metrics has consistently improved.
}
\label{fig:filtered}
\end{figure*}

\subsection{Preliminary Analysis of Training Data}

Prior investigations~\cite{zhangPerformancePlateausComprehensive2024,gonzalezEffectTrainingDataset2024a} have demonstrated that increasing data amount beyond a certain point yields diminishing performance improvements.
We look into the scalability limitations observed in large-scale SE training, particularly within the URGENT2025 challenge~\cite{koheiurgent2025}.
URGENT25 applied a simple DNSMOS~\cite{reddyDnsmosNonIntrusivePerceptual2021} and voice-activity-detection-based filtering on some large-scale subsets of \textit{Track2}, and then did random selection to obtain the Track1 training data.
However, the larger dataset (i.e., the 60k-hour \textit{Track2}) failed to consistently outperform the smaller, simply curated subset (i.e., the 2,500-hour \textit{Track1} data) \cite{koheiurgent2025}, prompting a critical re-examination of the training data curation. 

We conducted a preliminary analysis of the URGENT2025 \textit{Track1} training set (2,500 hours), which comprises several popular datasets for speech enhancement, as shown in Table~\ref{tab:dataset}.
Consistent with the known quality variations in this challenge's data~\cite{Lessons-Zhang2025}, our analysis observed that some segments labeled as ``clean'' exhibit lower speech quality.
Figure~\ref{fig:badcase} shows some of the most common distortions that we found in the target ``clean'' speech of the training set:
\begin{itemize}
    \item \textbf{Wideband noise}. The speech signal is contaminated by some low-energy wideband noise during the recording process, as shown in Figure~\ref{fig:badcase}.
    \item \textbf{Narrowband noise}. We found that some of the target ``clean'' speech signals contain high-energy narrowband noise at infrasound frequencies. This case is usually found in some samples of the DNS5 LibriVox~\cite{dubeyICASSP2023Deep2024}, VCTK~\cite{veauxVoiceBankCorpus2013}, and EARS datasets~\cite{richterEARSAnechoicFullband2024}.
    \item \textbf{Instantaneous noise}. Some speech sources include the clicking sound of a button or the sound of hitting an object. For example, as Figure~\ref{fig:badcase} shows, in some of the speech sources in VCTK~\cite{veauxVoiceBankCorpus2013}, a relatively distinct button sound can be heard near the end of the recording. 
    \item \textbf{High-frequency distortion}. It is usually found in the 48 kHz Common Voice~\cite{ardilaCommonVoiceMassivelyMultilingual2020}. This may be attributed to the lossy MP3 encoding format adopted by Common Voice.
    \item \textbf{Multi-talker speech}. Some source signals contain the speech from multiple talkers, as found in DNS5 LibriVox~\cite{dubeyICASSP2023Deep2024}. 
    Since most speech enhancement systems are typically trained only with single-talker speech by default, the impact of this case on training the speech enhancement model is uncertain. It may be more detrimental to the training of generative methods than to that of discriminative methods.
\end{itemize}

\subsection{Data Curation Strategy}
\label{sec:data_curation}
To mitigate the detrimental effects of low-quality ``clean'' targets and test our hypothesis that data quality is more important than quantity for scaling SE models, we developed a data curation pipeline focused on identifying and selecting high-fidelity utterances from the URGENT2025 \textit{Track1} training set.

Our core strategy leverages non-intrusive speech quality and intelligibility metrics. Unlike intrusive metrics (e.g., SDR~\cite{fevotteBSS_EVALToolboxUser2005}, PESQ~\cite{rixPerceptualEvaluationSpeech2001}), which require a ground-truth reference, non-intrusive metrics predict quality or intelligibility solely from the single audio signal itself. This is essential for our purpose, as we need to assess the quality of the supposedly clean target speech files without access to a higher-quality reference.

\begin{table}[tb]
\centering
\caption{Quality Thresholds for Threshold-based Filtering.}
\label{tab:thresholds}
\begin{tabular}{lccc}
\toprule
\textbf{Metric} & \textbf{EARS\cite{richterEARSAnechoicFullband2024}} & \textbf{Common Voice(ZH)\cite{ardilaCommonVoiceMassivelyMultilingual2020}} & \textbf{Other} \\
\midrule
DNSMOS & 2.5 & 3.0 & 3.0 \\
SigMOS & 2.5 & 3.0 & 3.0 \\
UTMOS & 2.5 & 3.0 & 3.0 \\
NISQA & 3.0 & 4.0 & 4.0 \\
SQUIM\_SDR & 0.0 & 0.0 & 20 \\
\bottomrule
\end{tabular}
\end{table}

\begin{table}[tb]
\caption{The durations (in hours) of speech source in each subset of URGENT2025 \textit{Track1}. The table lists the original and retained durations before and after filtering, respectively.}
\label{tab:dataset}
\centering
\begin{tabular}{@{}l|cc@{}}
\toprule
\multicolumn{1}{c|}{\textbf{Dataset}} &
  \textbf{\begin{tabular}[c]{@{}c@{}}Duration of\\ Full Dataset\end{tabular}} &
  \textbf{\begin{tabular}[c]{@{}c@{}}Duration of\\ Filtered dataset\end{tabular}} \\ \midrule
LibriVox from DNS5 \cite{dubeyICASSP2023Deep2024} & $\sim$350  & $\sim$150 \\
LibriTTS \cite{zenLibriTTSCorpusDerived2019}           & $\sim$200  & $\sim$109 \\
VCTK \cite{veauxVoiceBankCorpus2013}           & $\sim$80   & $\sim$44  \\
EARS \cite{richterEARSAnechoicFullband2024}            & $\sim$107  & $\sim$16  \\
MLS-HQ \cite{pratapMLSLargeScaleMultilingual2020,koheiurgent2025}            & $\sim$450  & $\sim$129 \\
CommonVoice 19.0 \cite{ardilaCommonVoiceMassivelyMultilingual2020}  & $\sim$1300 & $\sim$250 \\
WSJ \cite{garofolojohns.CSRIWSJ0Complete2007}             & $\sim$85   & not used  \\ \midrule
In Total           & $\sim$2500 & $\sim$700 \\ \bottomrule
\end{tabular}
\end{table}

Specifically, we employed the following widely used DNN-based non-intrusive metrics to score each ``clean'' speech utterance in the 2,500-hour corpus:
\begin{itemize}
\item \textbf{DNSMOS}~\cite{reddyDnsmosNonIntrusivePerceptual2021}: Predicts perceptual quality (MOS) via separate SigMOS (speech distortion) and NoiseMOS (background noise) scores, specifically optimized for noisy speech conditions.
\item \textbf{NISQA}~\cite{mittagNISQADeepCNNSelfAttention2021}: Estimates overall speech quality (MOS) using a CNN-SelfAttention architecture, capturing both distortion types and speech discontinuity impacts.
\item \textbf{SIGMOS}~\cite{risteaICASSP2024Speech2025}: Estimates the P.804 audio quality dimensions score non-intrusively, focusing on mimicking human perception of audio quality.
\item \textbf{Torchaudio-SQUIM-SDR}~\cite{kumarTorchaudioSquimReferenceLessSpeech2023}: SI-SDR~\cite{rouxSDRHalfbakedWell2019} estimated by Torchaudio-SQUIM, a multi-task model predicting three key intrusive metrics (STOI~\cite{taalAlgorithmIntelligibilityPrediction2011}, PESQ~\cite{rixPerceptualEvaluationSpeech2001}, SI-SDR~\cite{rouxSDRHalfbakedWell2019} ) simultaneously from a single waveform.
\item \textbf{UTMOS}~\cite{saekiUTMOSUTokyoSaruLabSystem2022}: A transformer-based model predicting MOS, robust to diverse languages and acoustic conditions, leveraging large-scale pretraining.
\end{itemize}

To address the issue of suboptimal ``clean'' targets, we implemented a two-stage data curation pipeline for the URGENT2025 \textit{Track1} training set:

\subsubsection{Threshold-Based Filtering (TBF)} We applied dataset-specific minimum quality thresholds using multiple non-intrusive metrics. Utterances failing to meet any threshold for their respective dataset were excluded. 
Due to the inherent quality differences across subset data sources, we applied different thresholds specific to each subset. The specific thresholds used for each dataset type and metric are detailed in Table~\ref{tab:thresholds}. This initial filtering step selects a curated subset of approximately 700 hours of higher-quality ``clean'' speech.
Dataset quantitative changes after TBF are reported in Table~\ref{tab:dataset}, while Figure~\ref{fig:filtered} illustrates the distribution of each metric before and after TBF.
We also report three additional metrics not used in the TBF, which are DNSMOS Pro \cite{cumlinDNSMOSProReducedSize2024}, VQScore \cite{fuSelfSupervisedSpeechQuality2023}, and Distill-MOS \cite{stahlDistillationPruningScalable2025}. 
It can be observed that the distribution of all the metrics has improved significantly and consistently for both used and unused metrics in the TBF.

\subsubsection{Quality Ranking and Subset Selection}
Within the ~700-hour TBF dataset, we further refined the selection of data based on the quality ranking. For each utterance, the scores from all five metrics are normalized. 
Normalization is performed by subtracting the mean and then dividing by the standard deviation calculated over the entire 700-hour TBF dataset for each metric individually.
The normalized scores for each utterance are summed to an overall score. The speech sources are ranked by their overall score. We selected the top-ranked 100-hour, 350-hour, and 700-hour subsets (including the 700-hour TBF dataset itself) to create three high-quality subsets, ranging from small-scale to large-scale.
To isolate the impact of our quality-centric curation strategy, we generated uniformly random subsets of identical sizes (100h, 350h, 700h) from the original, unfiltered 2,500-hour training set. This experimental design enabled us to rigorously investigate whether prioritizing data quality through targeted filtering and selection offers greater performance benefits than simply increasing data quantity, and to determine if smaller, high-quality datasets can outperform much larger, uncurated ones.

\section{Experiments}
\label{sec:exp_results}

\begin{table*}[thb]
\setlength{\tabcolsep}{4pt}
\caption{Results Comparison on 2500-hour full training set and Threshold-Based Filtered (TBF) 700-hour training set. DNSMOS Pro and DitillMOS are not used in the threshold-based filtering. Except for LSD, the larger numbers are better for other metrics.}
\label{tab:exp_tbf}
\centering
\begin{tabular}{@{}cccccccc|cccc@{}}
\toprule
\multirow{2}{*}{Model} &
  \multicolumn{1}{c}{\multirow{2}{*}{\begin{tabular}[c]{@{}c@{}}Training \\ source\end{tabular}}} &
  \multicolumn{6}{c}{non-intrusive metrics} &
  \multicolumn{4}{|c}{intrusive metrics} \\ \cmidrule(l){3-12} 
 &
  \multicolumn{1}{c}{} &
  DNSMOS & NISQA & UTMOS & SIGMOS & DNSMOS Pro* & DistillMOS* & SDR & PESQ & ESTOI & LSD $\downarrow$ \\ \midrule
\begin{tabular}[c]{@{}c@{}}Urgent2 Baseline \cite{koheiurgent2025} \\ (TF-GridNet \cite{wangTFGridNetMakingTimeFrequency2023})\end{tabular} &
  2500h-full & 2.85 & 2.77 & 1.92 & 2.61 & 3.39 & 3.19 & 10.24 & 2.24 & 0.76 & 2.72 \\ \midrule
BSRNN & 2500h-full & 2.80 & 2.83 & 2.00 &
  2.61 & 3.22 & 3.13 & 10.89 & 2.39 & 0.79 & 2.99 \\
BSRNN & 700h-TBF & \textbf{2.85} & 2.96 & 2.05 & \textbf{2.76} & \textbf{3.43} & 3.16 & 10.71 & 2.36 & 0.78 & 2.89 \\
BSRNN$_{init}$ & 700h-TBF & 2.84 & \textbf{2.98} & \textbf{2.06} & 2.74 & 3.41 & \textbf{3.21} & \textbf{10.98} & \textbf{2.41} & \textbf{0.79} & \textbf{2.87} \\ \midrule
BSRNN-Flow & 2500h-full & 2.84 & 3.01 & 1.99 & 2.88 & 3.56 & 3.32 & \textbf{9.20} & \textbf{2.05} & \textbf{0.74} & \textbf{3.79} \\
BSRNN-Flow & 700h-TBF & \textbf{2.90} & \textbf{3.07} & \textbf{2.04} & \textbf{3.01} & \textbf{3.87} & \textbf{3.44} & 8.95 & 2.02 & 0.73 & 3.88 \\
 \bottomrule
\end{tabular}
\end{table*}

\subsection{Datasets and Evaluation}

We adapt the \textit{Track1} training data from the URGENT2025 challenge \cite{koheiurgent2025} to train our models.
Its speech source consists of several commonly used speech enhancement datasets, which are listed in Table \ref{tab:dataset}. 
The noise source and room impulse response are identical to those in the URGENT2025 \textit{Track1} training data.
The dynamic simulation strategy\footnote{The training data and simulation scripts can be found at \url{https://github.com/urgent-challenge/urgent2025_challenge}} is used to train all models in this paper. We apply an additional high-pass filter to the clean speech source before simulation, aiming to remove potential low-frequency, narrow-band noise at infrasound frequencies. The cutoff frequency of the high-pass filter is $75$ Hz.

The blind test set of URGENT 2025 is applied to evaluate the models. We report six non-intrusive metrics, including DNSMOS \cite{reddyDnsmosNonIntrusivePerceptual2021}, NISQA~\cite{mittagNISQADeepCNNSelfAttention2021}, UTMOS~\cite{saekiUTMOSUTokyoSaruLabSystem2022}, SIGMOS~\cite{risteaICASSP2024Speech2025},  DNSMOS Pro \cite{cumlinDNSMOSProReducedSize2024}, and Distill-MOS \cite{stahlDistillationPruningScalable2025} in the evaluation, where the former four are used in the TBF, while the other two are not.
For intrusive metrics, we reported the signal-to-distortion ratio (SDR)~\cite{vincentPerformanceMeasurementBlind2006}, perceptual evaluation of speech quality (PESQ)~\cite{rixPerceptualEvaluationSpeech2001}, extended short-time objective intelligibility (ESTOI)~\cite{rixPerceptualEvaluationSpeech2001}, and log-spectral distance (LSD)~\cite{grayDistanceMeasuresSpeech1976}.

\subsection{Speech Enhancement Models}

The training and evaluation data in URGENT 2025 have various sampling frequencies (SF) of \{8, 16, 22.05, 24, 32, 44.1, 48\} kHz, which require the SE model to be able to handle multiple SF.  

\subsubsection{Discriminative Models} We adopt the Band-Split RNN (BSRNN) \cite{yuEfficientMonauralSpeech2023} as the discriminative SE model in this paper.
BSRNN has demonstrated its superior performance in recent source separation \cite{luoMusicSourceSeparation2023} and SE works \cite{chenComplexityScalingSpeech2024}.
It estimates the enhanced speech in the time-frequency (TF) domain.
BSRNN follows the ideal of the state-of-the-art dual-path design \cite{yangTFPSNetTimeFrequencyDomain2022,wangTFGridNetMakingTimeFrequency2023} in the TF domain, which typically applies two kinds of sequence modeling modules alternately to the frequency feature and the time feature in the TF domain.
The BSRNN model supports inputs with different SF and reduces the frequency dimension by splitting the frequency bins into a hand-crafted band split, effectively balancing computational efficiency with performance.
We use the BSRNN implementation from the ESPnet-SE~\cite{liESPnetSEEndEndSpeech2021} toolkit, and an L1-based time-domain plus frequency-domain multiresolution loss \cite{luLowDistortionMultiChannelSpeech2022} is used to train the discriminative BSRNN SE models. All discriminative BSRNN SE models have a parameter size of $38$ million.

\subsubsection{Generative Models} 
We follow a recent work named FlowSE~\cite{leeFlowSEFlowMatchingbased2025} to build generative SE models. It extends the flow matching method~\cite{lipmanFlowMatchingGenerative2023} to a conditional flow matching model that generates clean speech conditioned by the noisy speech.
The original FlowSE applies the Noise Conditional Score Network (NCSN++) \cite{songScoreBasedGenerativeModeling2021} as the backbone model. 
Since our dataset comprises speech of different sampling frequencies, we reimplement an improved BSRNN as a replacement for the NCSN++ to estimate the conditional vector field.  In the remainder of this paper, we refer to the BSRNN-based flow matching SE as BSRNN-Flow.
All generative BSRNN-Flow SE models have a parameter size of $103$ million.

\begin{figure*}[!htbp]
\centerline{\includegraphics[width=0.95\linewidth]{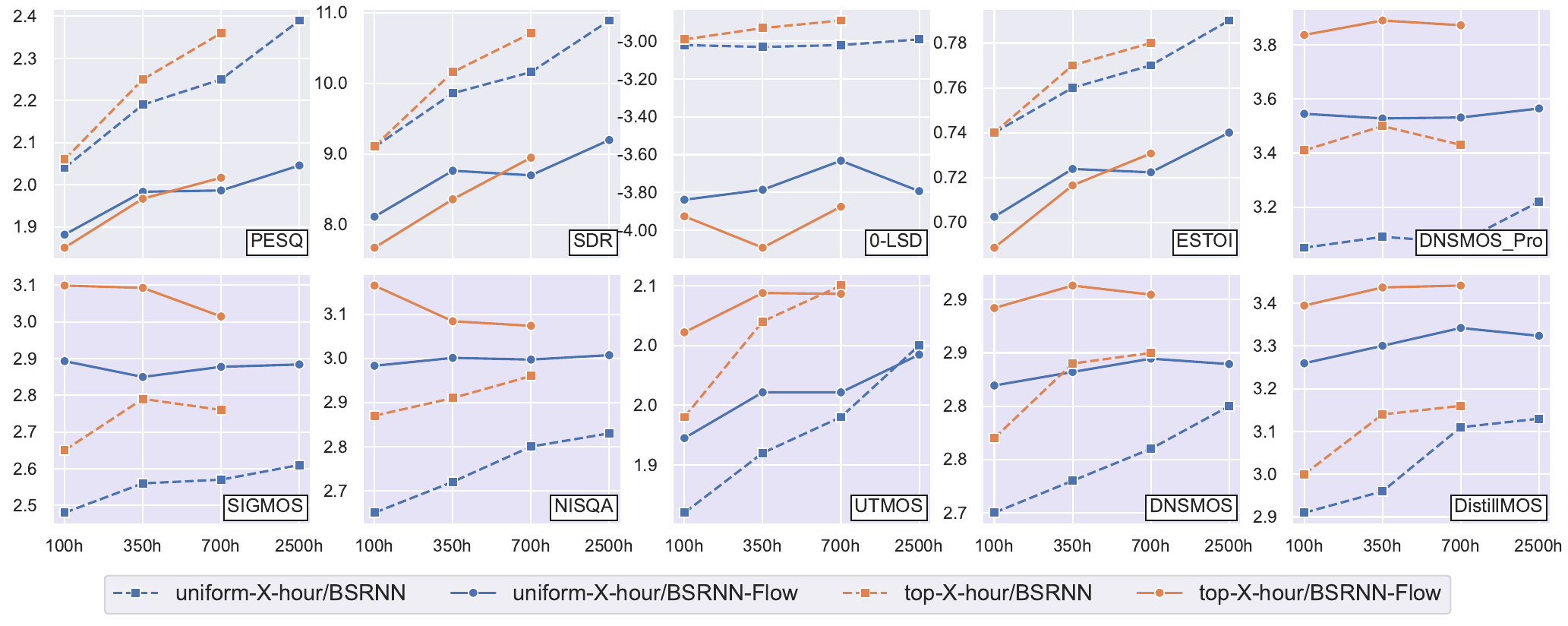}}
\caption{Impact of training data selection and scale on SE performance. Both discriminative BSRNN and generative BSRNN-Flow are evaluated. The X-axis is the amount of training speech source. The orange lines mark the systems trained with selected top-X-hour speech source; the blue lines mark the systems trained with uniformly sampled X-hour speech source. Intrusive and non-intrusive metrics are represented in different background colors.}
\label{fig:scaling_source}
\end{figure*}

\subsection{Less is More: Experimental Results in TBF}
We apply the threshold-based filtering (TBF) introduced in Section~\ref{sec:data_curation} to the 2,500-hour speech sources of the URGENT2025 \textit{Track1} training data (\textit{2500h-full}).
And then, as Figure~\ref{fig:filtered} suggests, a 700-hour subset with a better metric distribution is obtained, which we refer to as \textit{700h-TBF}.
We first train a discriminative SE baseline BSRNN on \textit{ 2500h-full}, and compare it with the TF-GridNet~\cite{wangTFGridNetMakingTimeFrequency2023} official baseline in URGENT2025.
The results in Table~\ref{tab:exp_tbf} show that our \textit{2500-full} baseline BSRNN achieves a performance comparable to the official baseline.

We train discriminative BSRNN and generative BSRNN-Flow on both \textit{2500-full} and \textit{700h-TBF} datasets.
Furthermore, we also train a discriminative BSRNN$_{init}$, which is trained on the \textit{700h-TBF} dataset with initialization from the model trained on \textit{2500-full}.
Comparing the results in Table~\ref{tab:exp_tbf}, there are two findings:
First, \textbf{data quality matters more than quantity}.
Both the BSRNN SE and BSRNN-Flow SE, trained on only \textit{700h-TBF}, consistently outperform their \textit{2500h-full} counterparts across all non-intrusive metrics, achieving on-par performance on intrusive metrics.
This indicates that training on a \textit{smaller and higher-quality subset} yields better generalization to perceptual quality metrics than using the larger, unprocessed data.
Second, the initialization provides an approach for the more efficient use of large-scale data. When initializing from the \textit{2500h-full} checkpoint, the \textit{700h-TBF}-trained model BSRNN$_{init}$ achieves best performance on multiple intrusive and non-intrusive metrics.
This suggests that warm-starting from a model pre-trained with large-scale data can extract maximal value from TBF data.

\subsection{Revisit Scaling in SE with Quality-Ranked Speech Source}

As introduced in Section~\ref{sec:data_curation}, we select the top-ranked 100-hour, 350-hour, and 700-hour (identical to \textit{700h-TBF}) subsets from the \textit{700h-TBF} training set.  
To isolate the impact of our quality-centric curation strategy and make a controlled comparison, we apply uniform random sampling to the \textit{2500h-full} training set, obtaining a series of subsets with the same quantity.
Figure~\ref{fig:scaling_source} makes a comprehensive comparison of both discriminative BSRNN and generative BSRNN-Flow with different quantities of training speech source.

 \begin{figure*}[htbp]
\centerline{\includegraphics[width=\linewidth]{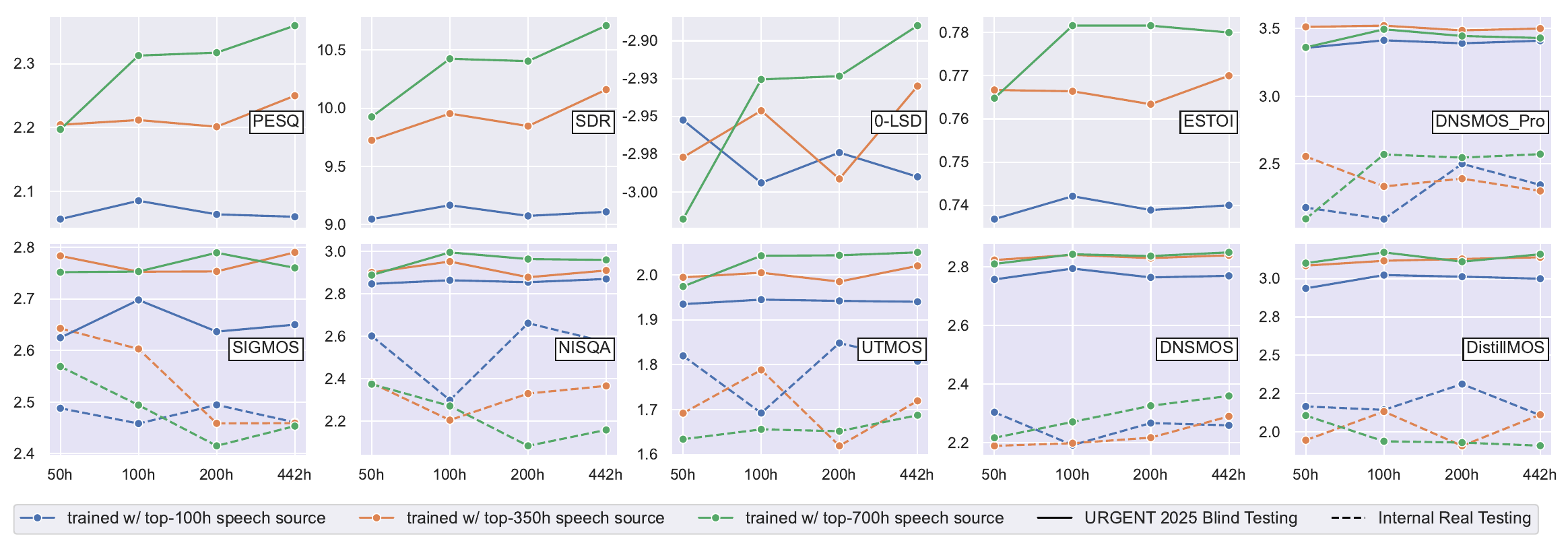}}
\caption{Impact of noise source scale in the training data. The X-axis is the amount of training noise source. Intrusive and non-intrusive metrics are represented in different background colors. We only report the non-intrusive metrics for the internal real testing, as the reference audios are not available.}
\label{fig:scaling_noise}
\end{figure*}

The most intuitive observation of the results in Figure~\ref{fig:scaling_source} is that the top-ranked subsets often surpass their randomly sampled counterparts. The discriminative BSRNN benefits more than the generative BSRNN-Flow.
For all the metrics, the discriminative BSRNNs trained on selected top-X-hour subsets outperform those trained on the same quantity of randomly sampled subsets. 
The generative BSRNN-Flow trained on selected top-X-hour subsets exhibits better performance than its counterparts trained on a randomly sampled subset across all non-intrusive metrics.
The quality ranking method does not show obvious differences in the intrusive metrics for generative methods.
This may be due to the inherent nature of generative models, that the generated signal can be regarded as drawn from a probability distribution, and is usually not well-aligned with the reference signal.

Another finding is that the performance gain of the quality ranking is more obvious on non-intrusive metrics for both the discriminative and generative models than on intrusive metrics. 
It is worth noting that, on some of the non-intrusive evaluation (e.g., DNSMOS Pro, SIGMOS, NISQA), the SE models trained on top-100-hour are even better than the \textit{2500-full} training set.

Furthermore, we found that the discriminative method scales better in terms of data size compared to the generative method. This conclusion is aligned with a previous study~\cite{gonzalezEffectTrainingDataset2024a}.
However, in \cite{gonzalezEffectTrainingDataset2024a}, the author reported diminishing returns for diffusion-based generative SE systems beyond 10 training hours on standard intrusive metrics (PESQ, ESTOI, and signal-to-noise ratio), while we can still observe a noticeable performance gain on PESQ, SDR, and ESTOI from 100-hour to 2500-hour data.
Differences in conclusions may arise from inconsistencies in the datasets and variations in the generative models.

\subsection{Scaling Investigation of the Amount of Noise Source}

We further conduct experiments on scaling the amount of noise source in the dynamic simulation. 
We uniformly sampled 50-hour, 100-hour, and 200-hour noise sources from the original URGRENT2025 442-hour additive noise sources, and performed the dynamic simulation with the top-ranked 100-hour, 350-hour, and 700-hour speech sources, respectively.
The experiments are conducted using discriminative BSRNN models and evaluated on the URGENT2025 blind testing set and another internal real testing set.
The internal testing set contains noisy speech recorded in various environments and can be regarded as an open test set.
The experimental results detailed in 10 evaluation metrics are presented in Figure~\ref{fig:scaling_noise}.
Surprisingly, we did not observe a pronounced scaling curve in either the URGENT2025 blind test set or the internal real test set.

This observed noise-scaling limitation may stem from the extensive diversity of noise types included within the URGENT2025 dataset. Our current scaling experiments focused exclusively on quantity via uniform random sampling, without accounting for potential scaling effects related to noise type richness~\cite{zhang2024scale}. Future work should systematically investigate this dimension, including the curation of noise sources, to gain a deeper understanding of this aspect. 

\section{Conclusion}
\label{sec:conclusion}
This paper revisits the diminishing marginal returns observed in speech enhancement (SE) performance when simply scaling the size of training data to massive scales.
Through a systematic analysis of the URGENT2025 dataset, we introduce a threshold-based filtering approach that leverages multiple non-intrusive quality metrics. 
Experiments on discriminative and generative SE models confirm that quality-curated subsets consistently outperform uniformly random datasets of identical size. Crucially, our most significant finding reveals that a rigorously selected 700-hour subset surpasses models trained on the full 2,500-hour dataset across multiple evaluation metrics. This discovery enlightens us that \textbf{data curation matters in scaling speech enhancement}.

The future works include:
1) Applying the proposed data curation methods to ultra-large datasets like the 60k-hour URGENT2025 \textit{Track2} to verify scalability limits.
2) Conducting targeted experiments on noise scaling dynamics, particularly how noise diversity (rather than quantity) affects enhancement performance.
It is hoped that this work will inspire researchers to explore more effective methods for utilizing suboptimal large-scale data to enhance the performance of SE.


\bibliographystyle{ieeetr}

\bibliography{refs}

\begin{thebibliography}{10}

\bibitem{loizouSpeechEnhancementTheory2007}
P.~C. Loizou, {\em Speech {{Enhancement}}: {{Theory}} and {{Practice}}}.
\newblock CRC Press, 2007.

\bibitem{xuRegressionApproachSpeech2015b}
Y.~Xu, J.~Du, L.-R. Dai, and C.-H. Lee, ``A {{Regression Approach}} to {{Speech Enhancement Based}} on {{Deep Neural Networks}},'' {\em IEEE/ACM Transactions on Audio, Speech, and Language Processing}, vol.~23, pp.~7--19, Jan. 2015.

\bibitem{wangSupervisedSpeechSeparation2018a}
D.~Wang and J.~Chen, ``Supervised {{Speech Separation Based}} on {{Deep Learning}}: {{An Overview}},'' {\em IEEE/ACM Transactions on Audio, Speech, and Language Processing}, vol.~26, pp.~1702--1726, Oct. 2018.

\bibitem{luoConvTasNetSurpassingIdeal2019}
Y.~Luo and N.~Mesgarani, ``Conv-{{TasNet}}: {{Surpassing Ideal Time}}--{{Frequency Magnitude Masking}} for {{Speech Separation}},'' {\em IEEE/ACM Transactions on Audio, Speech, and Language Processing}, vol.~27, pp.~1256--1266, Aug. 2019.

\bibitem{huDCCRNDeepComplex2020}
Y.~Hu, Y.~Liu, S.~Lv, M.~Xing, S.~Zhang, Y.~Fu, J.~Wu, B.~Zhang, and L.~Xie, ``{{DCCRN}}: {{Deep Complex Convolution Recurrent Network}} for {{Phase-Aware Speech Enhancement}},'' in {\em Interspeech 2020}, pp.~2472--2476, ISCA, 2020.

\bibitem{dubeyICASSP2023Deep2024}
H.~Dubey, A.~Aazami, V.~Gopal, B.~Naderi, S.~Braun, R.~Cutler, A.~Ju, M.~Zohourian, M.~Tang, M.~Golestaneh, and R.~Aichner, ``{{ICASSP}} 2023 {{Deep Noise Suppression Challenge}},'' {\em IEEE Open Journal of Signal Processing}, vol.~5, pp.~725--737, 2024.

\bibitem{uni-versa-ext}
W.~Wang, W.~Zhang, C.~Li, J.~Shi, S.~Watanabe, and Y.~Qian, ``Improving speech enhancement with multi-metric supervision from learned quality assessment,'' {\em arXiv preprint arXiv:2506.12260}, 2025.

\bibitem{luConditionalDiffusionProbabilistic2022}
Y.-J. Lu, Z.-Q. Wang, S.~Watanabe, A.~Richard, C.~Yu, and Y.~Tsao, ``Conditional {{Diffusion Probabilistic Model}} for {{Speech Enhancement}},'' in {\em {{ICASSP}} 2022 - 2022 {{IEEE International Conference}} on {{Acoustics}}, {{Speech}} and {{Signal Processing}} ({{ICASSP}})}, pp.~7402--7406, 2022.

\bibitem{liDiffusionbasedGenerativeModeling2024}
C.~Li, S.~Cornell, S.~Watanabe, and Y.~Qian, ``Diffusion-{{Based Generative Modeling With Discriminative Guidance}} for {{Streamable Speech Enhancement}},'' in {\em 2024 {{IEEE Spoken Language Technology Workshop}} ({{SLT}})}, pp.~333--340, Dec. 2024.

\bibitem{richterSpeechEnhancementDereverberation2023}
J.~Richter, S.~Welker, J.-M. Lemercier, B.~Lay, and T.~Gerkmann, ``Speech {{Enhancement}} and {{Dereverberation With Diffusion-Based Generative Models}},'' {\em IEEE/ACM Transactions on Audio, Speech, and Language Processing}, vol.~31, pp.~2351--2364, 2023.

\bibitem{leeFlowSEFlowMatchingbased2025}
S.~Lee, S.~Cheong, S.~Han, and J.~W. Shin, ``{{FlowSE}}: {{Flow Matching-based Speech Enhancement}},'' in {\em {{ICASSP}} 2025 - 2025 {{IEEE International Conference}} on {{Acoustics}}, {{Speech}} and {{Signal Processing}} ({{ICASSP}})}, pp.~1--5, Apr. 2025.

\bibitem{valentini-botinhaoSpeechEnhancementNoiseRobust2016}
C.~{Valentini-Botinhao}, X.~Wang, S.~Takaki, and J.~Yamagishi, ``Speech {{Enhancement}} for a {{Noise-Robust Text-to-Speech Synthesis System Using Deep Recurrent Neural Networks}},'' in {\em Proc. {{Interspeech}} 2016}, pp.~352--356, 2016.

\bibitem{zhangPerformancePlateausComprehensive2024}
W.~Zhang, K.~Saijo, J.-w. Jung, C.~Li, S.~Watanabe, and Y.~Qian, ``Beyond {{Performance Plateaus}}: {{A Comprehensive Study}} on {{Scalability}} in {{Speech Enhancement}},'' in {\em Proc. {{Interspeech}} 2024}, pp.~1740--1744, 2024.

\bibitem{gonzalezEffectTrainingDataset2024a}
P.~Gonzalez, Z.-H. Tan, J.~{\O}stergaard, J.~Jensen, T.~S. Alstr{\o}m, and T.~May, ``The {{Effect}} of {{Training Dataset Size}} on {{Discriminative}} and {{Diffusion-Based Speech Enhancement Systems}},'' {\em IEEE Signal Processing Letters}, vol.~31, pp.~2225--2229, 2024.

\bibitem{zhangURGENTChallengeUniversality2024}
W.~Zhang, R.~Scheibler, K.~Saijo, S.~Cornell, C.~Li, Z.~Ni, A.~Kumar, J.~Pirklbauer, M.~Sach, S.~Watanabe, T.~Fingscheidt, and Y.~Qian, ``{{URGENT Challenge}}: {{Universality}}, {{Robustness}}, and {{Generalizability For Speech Enhancement}},'' in {\em Interspeech 2024}, pp.~4868--4872, 2024.

\bibitem{koheiurgent2025}
K.~Saijo, W.~Zhang, S.~Cornell, R.~Scheibler, C.~Li, Z.~Ni, A.~Kumar, M.~Sach, Y.~Fu, W.~Wang, T.~Fingscheidt, and S.~Watanabe, ``{{Interspeech}} 2025 {{URGENT}} {{Speech}} {{Enhancement}} {{Challenge}},'' in {\em Interspeech 2026}, 2026.

\bibitem{kearns2014librivox}
J.~Kearns, ``Librivox: Free public domain audiobooks,'' {\em Reference Reviews}, vol.~28, no.~1, pp.~7--8, 2014.

\bibitem{ardilaCommonVoiceMassivelyMultilingual2020}
R.~Ardila, M.~Branson, K.~Davis, M.~Kohler, J.~Meyer, M.~Henretty, R.~Morais, L.~Saunders, F.~Tyers, and G.~Weber, ``Common {{Voice}}: {{A Massively-Multilingual Speech Corpus}},'' in {\em Proceedings of the {{Twelfth Language Resources}} and {{Evaluation Conference}}} (N.~Calzolari, F.~Béchet, P.~Blache, K.~Choukri, C.~Cieri, T.~Declerck, S.~Goggi, H.~Isahara, B.~Maegaard, J.~Mariani, H.~Mazo, A.~Moreno, J.~Odijk, and S.~Piperidis, eds.), pp.~4218--4222, European Language Resources Association, 2020.

\bibitem{veauxVoiceBankCorpus2013}
C.~Veaux, J.~Yamagishi, and S.~King, ``The voice bank corpus: {{Design}}, collection and data analysis of a large regional accent speech database,'' in {\em 2013 {{International Conference Oriental COCOSDA}} Held Jointly with 2013 {{Conference}} on {{Asian Spoken Language Research}} and {{Evaluation}} ({{O-COCOSDA}}/{{CASLRE}})}, pp.~1--4, 2013.

\bibitem{Lessons-Zhang2025}
W.~Zhang, K.~Saijo, S.~Cornell, R.~Scheibler, C.~Li, Z.~Ni, A.~Kumar, M.~Sach, W.~Wang, Y.~Fu, S.~Watanabe, T.~Fingscheidt, and Y.~Qian, ``Lessons learned from the {URGENT} 2024 speech enhancement challenge,'' {\em Accepted by Interspeech}, 2025.

\bibitem{cumlinDNSMOSProReducedSize2024}
F.~Cumlin, X.~Liang, V.~Ungureanu, C.~K.~A.~Reddy, C.~Sch{\"u}ldt, and S.~Chatterjee, ``{{DNSMOS Pro}}: {{A Reduced-Size DNN}} for {{Probabilistic MOS}} of {{Speech}},'' in {\em Proc. {{Interspeech}} 2024}, pp.~4818--4822, 2024.

\bibitem{fuSelfSupervisedSpeechQuality2023}
S.-W. Fu, K.-H. Hung, Y.~Tsao, and Y.-C.~F. Wang, ``Self-{{Supervised Speech Quality Estimation}} and {{Enhancement Using Only Clean Speech}},'' in {\em The {{Twelfth International Conference}} on {{Learning Representations}}}, Oct. 2023.

\bibitem{stahlDistillationPruningScalable2025}
B.~Stahl and H.~Gamper, ``Distillation and {{Pruning}} for {{Scalable Self-Supervised Representation-Based Speech Quality Assessment}},'' in {\em {{ICASSP}} 2025 - 2025 {{IEEE International Conference}} on {{Acoustics}}, {{Speech}} and {{Signal Processing}} ({{ICASSP}})}, pp.~1--5, Apr. 2025.

\bibitem{reddyDnsmosNonIntrusivePerceptual2021}
C.~K.~A. Reddy, V.~Gopal, and R.~Cutler, ``Dnsmos: {{A Non-Intrusive Perceptual Objective Speech Quality Metric}} to {{Evaluate Noise Suppressors}},'' in {\em {{ICASSP}} 2021 - 2021 {{IEEE International Conference}} on {{Acoustics}}, {{Speech}} and {{Signal Processing}} ({{ICASSP}})}, pp.~6493--6497, June 2021.

\bibitem{richterEARSAnechoicFullband2024}
J.~Richter, Y.-C. Wu, S.~Krenn, S.~Welker, B.~Lay, S.~Watanabe, A.~Richard, and T.~Gerkmann, ``{{EARS}}: {{An Anechoic Fullband Speech Dataset Benchmarked}} for {{Speech Enhancement}} and {{Dereverberation}},'' in {\em Proc. {{Interspeech}} 2024}, pp.~4873--4877, 2024.

\bibitem{fevotteBSS_EVALToolboxUser2005}
C.~F{\'e}votte, R.~Gribonval, and E.~Vincent, ``{{BSS}}\_{{EVAL Toolbox User Guide}} -- {{Revision}} 2.0,'' report, 2005.

\bibitem{rixPerceptualEvaluationSpeech2001}
A.~W. Rix, J.~G. Beerends, M.~P. Hollier, and A.~P. Hekstra, ``Perceptual evaluation of speech quality ({{PESQ}})-a new method for speech quality assessment of telephone networks and codecs,'' in {\em 2001 {{IEEE International Conference}} on {{Acoustics}}, {{Speech}}, and {{Signal Processing}}. {{Proceedings}} ({{Cat}}. {{No}}.{{01CH37221}})}, vol.~2, (Utah, USA), pp.~749--752 vol.2, May 2001.

\bibitem{zenLibriTTSCorpusDerived2019}
H.~Zen, V.~Dang, R.~Clark, Y.~Zhang, R.~J. Weiss, Y.~Jia, Z.~Chen, and Y.~Wu, ``{{LibriTTS}}: {{A Corpus Derived}} from {{LibriSpeech}} for {{Text-to-Speech}},'' in {\em Proc. {{Interspeech}} 2019}, pp.~1526--1530, 2019.

\bibitem{pratapMLSLargeScaleMultilingual2020}
V.~Pratap, Q.~Xu, A.~Sriram, G.~Synnaeve, and R.~Collobert, ``{{MLS}}: {{A Large-Scale Multilingual Dataset}} for {{Speech Research}},'' in {\em Proc. {{Interspeech}} 2020}, pp.~2757--2761, 2020.

\bibitem{garofolojohns.CSRIWSJ0Complete2007}
{Garofolo, John S.}, {Graff, David}, {Paul, Doug}, and {Pallett, David}, ``{{CSR-I}} ({{WSJ0}}) {{Complete}},'' 2007.

\bibitem{mittagNISQADeepCNNSelfAttention2021}
G.~Mittag, B.~Naderi, A.~Chehadi, and S.~M{\"o}ller, ``{{NISQA}}: {{A Deep CNN-Self-Attention Model}} for {{Multidimensional Speech Quality Prediction}} with {{Crowdsourced Datasets}},'' in {\em Proc. {{Interspeech}} 2021}, pp.~2127--2131, 2021.

\bibitem{risteaICASSP2024Speech2025}
N.-C. Ristea, B.~Naderi, A.~Saabas, R.~Cutler, S.~Braun, and S.~Branets, ``{{ICASSP}} 2024 {{Speech Signal Improvement Challenge}},'' {\em IEEE Open Journal of Signal Processing}, vol.~6, pp.~238--246, 2025.

\bibitem{kumarTorchaudioSquimReferenceLessSpeech2023}
A.~Kumar, K.~Tan, Z.~Ni, P.~Manocha, X.~Zhang, E.~Henderson, and B.~Xu, ``Torchaudio-{{Squim}}: {{Reference-Less Speech Quality}} and {{Intelligibility Measures}} in {{Torchaudio}},'' in {\em {{ICASSP}} 2023 - 2023 {{IEEE International Conference}} on {{Acoustics}}, {{Speech}} and {{Signal Processing}} ({{ICASSP}})}, pp.~1--5, June 2023.

\bibitem{rouxSDRHalfbakedWell2019}
J.~L. Roux, S.~Wisdom, H.~Erdogan, and J.~R. Hershey, ``{{SDR}} -- {{Half-baked}} or {{Well Done}}?,'' in {\em {{ICASSP}} 2019 - 2019 {{IEEE International Conference}} on {{Acoustics}}, {{Speech}} and {{Signal Processing}} ({{ICASSP}})}, pp.~626--630, May 2019.

\bibitem{taalAlgorithmIntelligibilityPrediction2011}
C.~H. Taal, R.~C. Hendriks, R.~Heusdens, and J.~Jensen, ``An {{Algorithm}} for {{Intelligibility Prediction}} of {{Time}}--{{Frequency Weighted Noisy Speech}},'' {\em IEEE Transactions on Audio, Speech, and Language Processing}, vol.~19, pp.~2125--2136, Sept. 2011.

\bibitem{saekiUTMOSUTokyoSaruLabSystem2022}
T.~Saeki, D.~Xin, W.~Nakata, T.~Koriyama, S.~Takamichi, and H.~Saruwatari, ``{{UTMOS}}: {{UTokyo-SaruLab System}} for {{VoiceMOS Challenge}} 2022,'' in {\em Proc. {{Interspeech}} 2022}, pp.~4521--4525, 2022.

\bibitem{wangTFGridNetMakingTimeFrequency2023}
Z.-Q. Wang, S.~Cornell, S.~Choi, Y.~Lee, B.-Y. Kim, and S.~Watanabe, ``{TF-GridNet}: Making time-frequency domain models great again for monaural speaker separation,'' in {\em {{ICASSP}} 2023 - 2023 {{IEEE International Conference}} on {{Acoustics}}, {{Speech}} and {{Signal Processing}} ({{ICASSP}})}, 2023.

\bibitem{vincentPerformanceMeasurementBlind2006}
E.~Vincent, R.~Gribonval, and C.~Fevotte, ``Performance measurement in blind audio source separation,'' {\em IEEE Transactions on Audio, Speech, and Language Processing}, vol.~14, pp.~1462--1469, July 2006.

\bibitem{grayDistanceMeasuresSpeech1976}
A.~Gray and J.~Markel, ``Distance measures for speech processing,'' {\em IEEE Transactions on Acoustics, Speech, and Signal Processing}, vol.~24, pp.~380--391, Oct. 1976.

\bibitem{yuEfficientMonauralSpeech2023}
J.~Yu and Y.~Luo, ``Efficient {{Monaural Speech Enhancement}} with {{Universal Sample Rate Band-Split RNN}},'' in {\em {{ICASSP}} 2023 - 2023 {{IEEE International Conference}} on {{Acoustics}}, {{Speech}} and {{Signal Processing}} ({{ICASSP}})}, pp.~1--5, June 2023.

\bibitem{luoMusicSourceSeparation2023}
Y.~Luo and J.~Yu, ``Music {{Source Separation With Band-Split RNN}},'' {\em IEEE/ACM Transactions on Audio, Speech, and Language Processing}, vol.~31, pp.~1893--1901, 2023.

\bibitem{chenComplexityScalingSpeech2024}
H.~Chen, J.~Yu, and C.~Weng, ``Complexity {{Scaling}} for {{Speech Denoising}},'' in {\em {{ICASSP}} 2024 - 2024 {{IEEE International Conference}} on {{Acoustics}}, {{Speech}} and {{Signal Processing}} ({{ICASSP}})}, pp.~12276--12280, 2024.

\bibitem{yangTFPSNetTimeFrequencyDomain2022}
L.~Yang, W.~Liu, and W.~Wang, ``{{TFPSNet}}: {{Time-Frequency Domain Path Scanning Network}} for {{Speech Separation}},'' in {\em {{ICASSP}} 2022 - 2022 {{IEEE International Conference}} on {{Acoustics}}, {{Speech}} and {{Signal Processing}} ({{ICASSP}})}, pp.~6842--6846, 2022.

\bibitem{liESPnetSEEndEndSpeech2021}
C.~Li, J.~Shi, W.~Zhang, A.~S. Subramanian, X.~Chang, N.~Kamo, M.~Hira, T.~Hayashi, C.~Boeddeker, Z.~Chen, and S.~Watanabe, ``{{ESPnet-SE}}: {{End-To-End Speech Enhancement}} and {{Separation Toolkit Designed}} for {{ASR Integration}},'' in {\em 2021 {{IEEE Spoken Language Technology Workshop}} ({{SLT}})}, pp.~785--792, Jan. 2021.

\bibitem{luLowDistortionMultiChannelSpeech2022}
Y.-J. Lu, S.~Cornell, X.~Chang, W.~Zhang, C.~Li, Z.~Ni, Z.-Q. Wang, and S.~Watanabe, ``Towards {{Low-Distortion Multi-Channel Speech Enhancement}}: {{The ESPNET-Se Submission}} to the {{L3DAS22 Challenge}},'' in {\em {{ICASSP}} 2022 - 2022 {{IEEE International Conference}} on {{Acoustics}}, {{Speech}} and {{Signal Processing}} ({{ICASSP}})}, pp.~9201--9205, May 2022.

\bibitem{lipmanFlowMatchingGenerative2023}
Y.~Lipman, R.~T.~Q. Chen, H.~{Ben-Hamu}, M.~Nickel, and M.~Le, ``Flow {{Matching}} for {{Generative Modeling}},'' in {\em The {{Eleventh International Conference}} on {{Learning Representations}}}, Sept. 2022.

\bibitem{songScoreBasedGenerativeModeling2021}
Y.~Song, J.~{Sohl-Dickstein}, D.~P. Kingma, A.~Kumar, S.~Ermon, and B.~Poole, ``Score-{{Based Generative Modeling}} through {{Stochastic Differential Equations}},'' in {\em 9th {{International Conference}} on {{Learning Representations}}, {{ICLR}} 2021, {{Virtual Event}}, {{Austria}}, {{May}} 3-7, 2021}, OpenReview.net, 2021.

\bibitem{zhang2024scale}
L.~Zhang, W.~Zhang, C.~Li, and Y.~Qian, ``Scale this, not that: Investigating key dataset attributes for efficient speech enhancement scaling,'' {\em arXiv preprint arXiv:2412.14890}, 2024.

\end{thebibliography}

\end{document}